\documentclass[preprint,12pt]{elsarticle}

\usepackage{lineno,hyperref}
\modulolinenumbers[5]

\newcounter{bla}

\journal{Optics Communications}

\usepackage[bbgreekl]{mathbbol}
\usepackage{amsfonts}
\usepackage{amsmath}
\usepackage{amssymb}
\usepackage{booktabs}
\usepackage{tabularx}
\usepackage{mathtools}
\usepackage{pgfplots}
\usepackage{listings}
\usepackage{geometry}
\usepackage{pdflscape}
\usepackage{enumerate}
\usepackage{esvect}
\usepackage[figuresright]{rotating}
\usepackage{footnote}
\usepackage{listings}
\usepackage{xcolor}
\usepackage{enumitem}
\usepackage{tocloft}
\usepackage{textcomp}
\usepackage{braket}
\usepackage{nicefrac}
\usepackage{nccmath}
\usepackage{bm}
\usepackage{import}
\usepackage{dpfloat}
\usepackage{subcaption}
\usepackage{float}
\usepackage{afterpage}
\usepackage{multirow}
\usepackage{siunitx}
\usepackage{graphicx}
\usepackage{pdfcomment}
\usepackage{tikz}
\definecolor{Darkgreen}{rgb}{0,0.4,0}
\usepackage[ruled]{algorithm2e}
\graphicspath{{Figs/}}
\DeclarePairedDelimiter\abs{\lvert}{\rvert}%
\definecolor{light-gray}{gray}{0.95}

\usepackage{graphicx}
\newcommand{\CC}{%
	{\settoheight{\dimen0}{C}C\kern-.05em \resizebox{!}{\dimen0}{\raisebox{\depth}{++}}}}
\newcommand{\CCC}{%
	{\settoheight{\dimen0}{C}C/C\kern-.05em \resizebox{!}{\dimen0}{\raisebox{\depth}{++}}}}
\newcommand{\CS}{%
	{\settoheight{\dimen0}{C}C\kern-.05em \resizebox{!}{\dimen0}{\raisebox{\depth}{\#}}}}

\begin{document}
	\suppressfloats 
	
	\begin{frontmatter}
		
		\title{Sympathetic quantisation - a new approach to hologram quantisation}
		
		\author[mymainaddress]{Peter J. Christopher\corref{mycorrespondingauthor}}
		\cortext[mycorrespondingauthor]{Corresponding author}
		\ead{pjc209@cam.ac.uk}
		\ead[url]{www.peterjchristopher.me.uk}
		
		\author{Ralf Mouthaan}
        
    \author{A. Mohamed Soliman}
		
		\author{Timothy D. Wilkinson}
		
		\address[mymainaddress]{Centre of Molecular Materials, Photonics and Electronics, University of Cambridge}
		
		\begin{abstract}
		
		Spatial light modulators can typically only modulate the phase or the amplitude of an incident wavefront, with only a limited number of discrete values available. This is often accounted for in computer-generated holography algorithms by setting hologram pixel values to the nearest achievable value during what is known as quantisation. Sympathetic quantisation is an alternative to this nearest-neighbour approach that takes into account the underlying diffraction relationships in order to obtain a significantly improved post-quantisation performance. The concept of sympathetic quantisation is introduced in this paper and a simple implementation, soft sympathetic quantisation, is presented which is shown to improve mean squared error and structural similarity index error metrics by 50\% for the considered case of single-transform algorithms.
		
		\end{abstract}
		
		
	\end{frontmatter}
    
    \section{Introduction}
    
    The mathematical connections between physical geometry and interference patterns are widely used in applications such as sonar \cite{Goldstein1998, Creuze2005, Marshall2010} and radar \cite{Deming1997, Zhang2013, Mast1994}. Many applications involve taking a measured interference pattern and reconstructing the physical geometry that caused it. Computer generated holography (CGH) works in reverse, taking a target geometry and attempting to find an appropriate aperture modulation function that can reproduce it. Best known in display applications, CGH sees widespread application in a range of applications from lithography~\cite{Purvis2014,Turberfield2000} and optical manipulation~\cite{Grieve2009,Grier06,Melville2003} to imaging~\cite{Frauel2006, Svoboda2013, Sheen2001} and displays~\cite{Yamada2018,Maimone2017}.
    
    CGH relies on wavefront-modulating devices, known as spatial light modulators (SLMs), which can be amplitude-modulating or phase-modulating. Typically two categories of phase modulating device may be considered, binary and multi-level devices. Binary devices, often based on ferroelectric liquid crystals~(LCs), are fast-switching but only capable of two phase modulation states, $0$ and $\pi$. Multi-level devices, often based on nematic LCs, offer much lower frame rates but up to 1024 modulation levels between $0$ and $2\pi$. CGH algorithms need to take into account the restricted behaviour of these devices, often by setting obtained hologram pixel values to the nearest available modulation state. In this paper, we set out to present a new concept in CGH generation that we are calling \textit{sympathetic quantisation} (SQ). SQ is a novel approach that exploits the underlying relationships of Fraunhofer and Fresnel diffraction in order to improve quantisation behaviour in CGH.
    
    We develop a single example of this approach called \textit{soft sympathetic quantisation} (SSQ) which is designed for use with single-iteration time-multiplexed algorithms. Two such algorithms, One-Step Phase-Retrieval (OSPR) \cite{cable200453} or Single-Transform Time-Multiplexed (STTM) \cite{STTM}, are introduced. We then show how the addition of SSQ can offer significant quality benefits at very low cost on generation time. Finally we discuss the implications of this research and draw conclusions.
    
    \section{Background}
    
\begin{figure}[tb]
        \centering
        {\includegraphics[trim={0 0.5cm 0 0},width=0.56\linewidth,page=1]{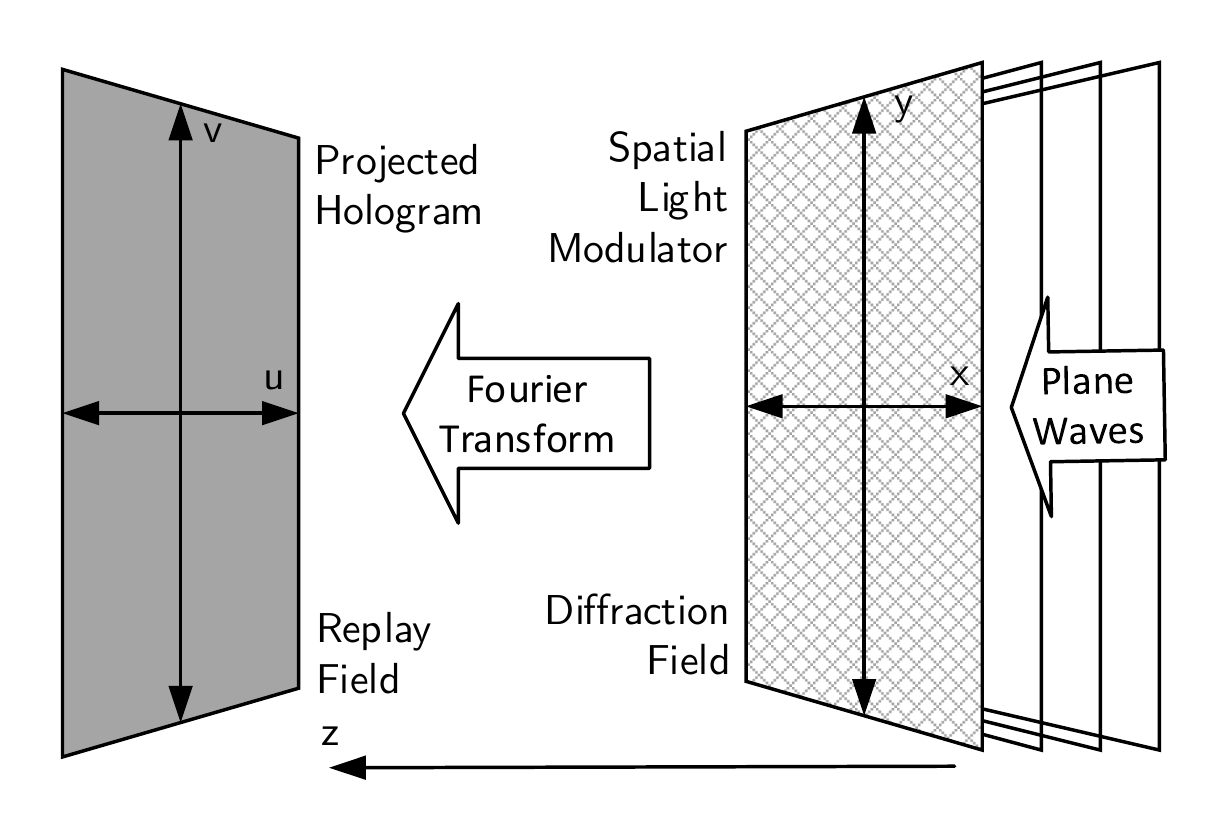}}
        {\includegraphics[trim={0 0 0 0},width=0.18\linewidth,page=1]{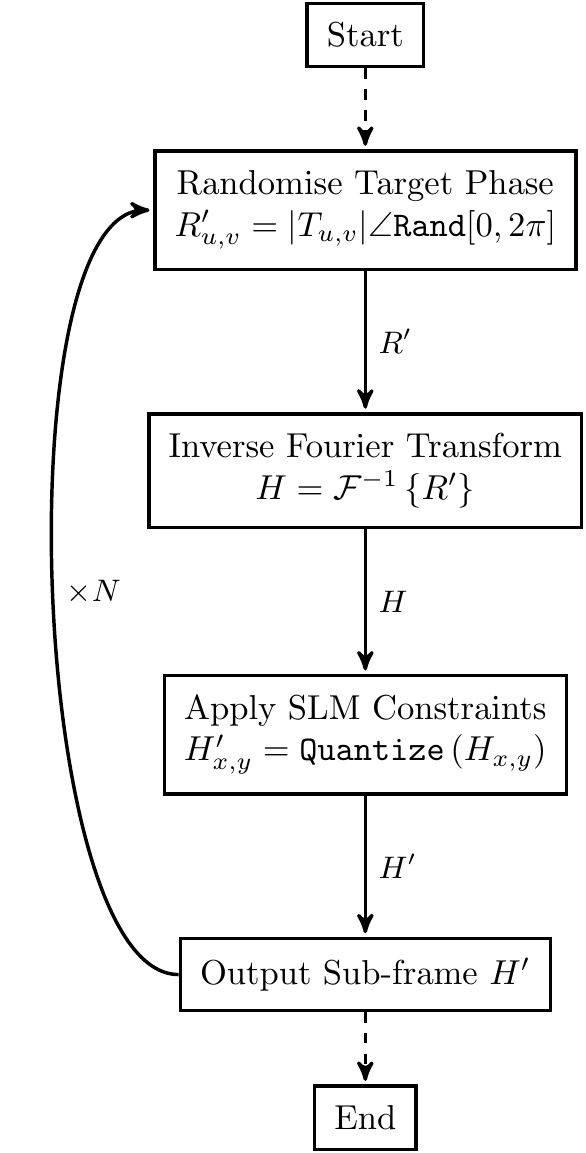}}
        {\includegraphics[trim={0 0 0 0},width=0.215\linewidth,page=1]{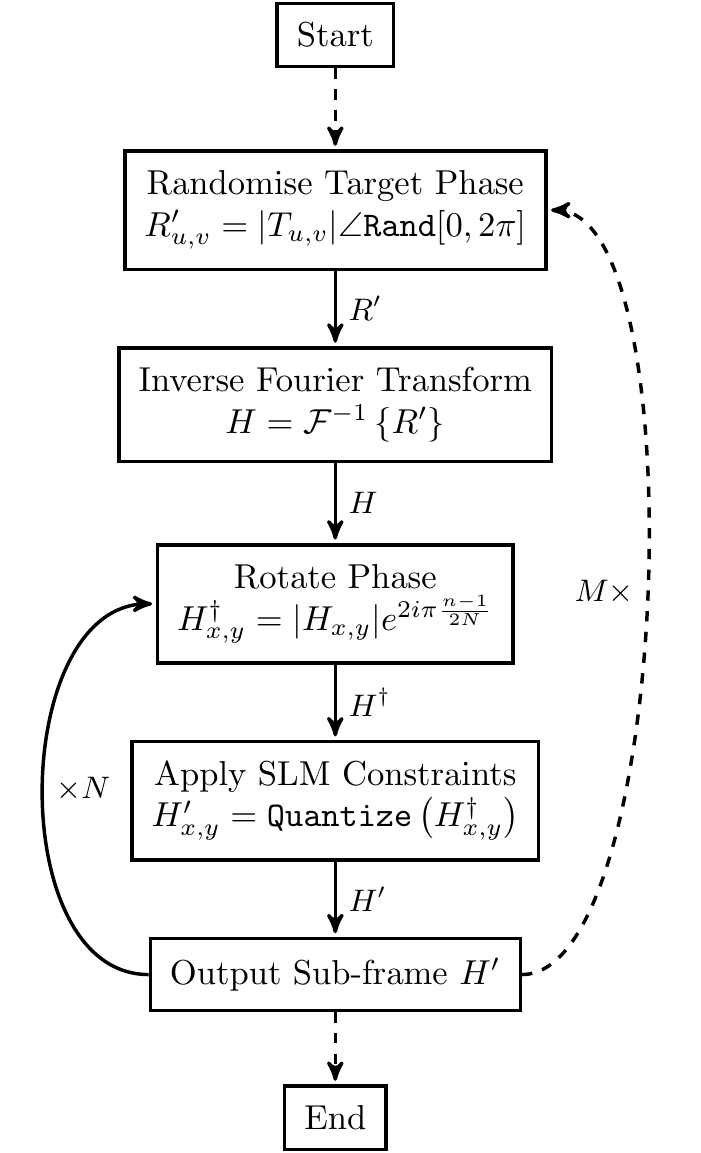}}
        \caption{Coordinate systems used in notation (left) with OSPR algorithm (left) and STTM algorithm (right). Centre and right figures used with permission from \cite{STTM}}
        \label{fig:algs}
    \end{figure}

    Mathematically, a hologram can be thought of as a Fourier transform with (Fresnel) or without (Fraunhofer) a quadratic phase term. Algorithms for generating holograms aim to find aperture function $H_{x,y}$ so that target image $T_{u,v} = \mathcal{F}\{H_{x,y}\}$, where $x$, $y$ are the spatial coordinates of the diffraction field, $u$ and $v$ are the spatial coordinates of the replay field and $\mathcal{F}$ represents the Fourier transform. The coordinates are shown in Figure~\ref{fig:algs} (left). In the case of discrete or pixellated data sets, $\mathcal{F}$ is calculated using the Discrete Fourier Transform (DFT). In a complex system there can be significant difference between the target image $T_{u,v}$ and actually generated replay field $R_{u,v}$. 
    
    For real-time visual applications, generation time is of great concern. In this context, time-multiplexed algorithms are used relying on human eye to time average many low-quality frames. Two such algorithms, One-Step Phase-Retrieval (OSPR) and Single-Transform Time-Multiplexed (STTM) are shown in Figure~\ref{fig:algs} (centre, right).
    
    The driving limitation of CGH generation is the modulation step where the aperture function is constrained by the limited modulation capabilities of the SLMs used \cite{HPS1}. For example, phase-modulating SLMs can only vary the phase of a pixel with the amplitude remaining unchanged. Often the modulation constraint is worsened by the digital nature of SLMs where the continuously modulated hologram is quantised to the discrete energy levels achievable by the device. For example, 8-bit phase only SLMs are constrained to a modulation angle steps of $\nicefrac{2\pi}{256}$ radians on the Argand circle. 
    
    The modulation and quantisation scheme used in these algorithms we have called nearest neighbour quantisation (NNQ). Here the ideal hologram value is changed to the closest achievable state in $\mathbb{C}$. If the pixels in the hologram can be assumed independent of each other then it can be shown that this is statistically the best procedure \cite{MostChangedPixel, STTM}. In this work we present an alternative approach that is aware of the correlated relationship between individual SLM pixels and uses this to improve the quality of single-iteration holograms.
    
    The mathematical form of the DFT allows us to write the following relationship between hologram $H_{x,y}$ and replay $R_{u,v}$ sampling points.
    
    \begin{equation} \label{DFT}
    R_{u,v} = \frac{1}{\sqrt{N_xN_y}} \sum_{N_x} \sum_{N_y} H_{x,y} \exp\left({\frac{ux}{N_x} + \frac{vy}{N_y}}\right)
    \end{equation}
    
    Naively this is $O(N_x^2N_y^2)$ but use of the Fast Fourier Transform (FFT) allows us to cut this down significantly to $O(N_xN_y \log{N_xN_y})$. Eq.~\ref{DFT} also allows us to write a relationship for the effect of a change in a hologram pixel $\Delta H_{x,y}$ on a given sampling point $\Delta R_{u,v}$ in the replay field.    
    
    \begin{equation} \label{changeEq}
    \Delta R_{u,v} = \frac{1}{\sqrt{N_xN_y}}\Delta H_{x,y} e^{\left[-2\pi i\left(\frac{ux}{N_x}+\frac{vy}{N_y}\right)\right]}
    \end{equation}
    
    There are several immediate observations here. The first is that every hologram pixel has an effect of equal magnitude at any given replay field pixel. It is only the summation of many contributions that leads to interference. A phase-only hologram is only capable of moving energy around in the replay field, not of creating or losing energy. 
    
    Secondly, Parseval's law or conservation of energy applies and accounts for the $\frac{1}{\sqrt{N_xN_y}}$ factor in Eqs.~(\ref{DFT}) \& (\ref{changeEq}). 
    
    \section{Sympathetic Quantisation Approach}
    
    The third observation for Eq.~(\ref{changeEq}) is the one that defines what we have termed sympathetic quantisation. This is the fact that modifying defined pairs of pixels synchronously can allow us limited control over the error caused by modulation or quantisation. For example if we take two pixels - $H_{x,y}$ and $H_{x\pm\frac{N_x}{2},y}$ - with identical $y$ coordinates and with $x$ coordinates separated by $\nicefrac{N_x}{2}$ we can localise 100\% of the error to 50\% of the columns with 50\% remaining error free. Analytically this can be seen from
    
    \begin{equation}
    \Delta R_{u,v} = \frac{1}{\sqrt{N_xN_y}}\left[
    \Delta H_{x,y} e^{\left[-2\pi i\left(\frac{ux}{N_x}+\frac{vy}{N_y}\right)\right]}+
    \Delta H_{x\pm\frac{N_x}{2},y} e^{\left[-2\pi i\left(\frac{u(x\pm\frac{N_x}{2})}{N_x}+\frac{vy}{N_y}\right)\right]}
    \right]
    \end{equation}
    
    which can be seen to cancel for values of $u=\pm2\pi$ provided $\Delta H_{x,y} = \Delta H_{x\pm\frac{N_x}{2},y}$. This result is of little practical use but, as we shall show later, depending on the location relationship between the pixels, the principle of sympathetic pixel quantisation can be exploited in a number of interesting ways.
    
    Two common features of hologram applications are relevant for this exploitation. Firstly, human vision is phase insensitive with the eye seeing the intensity of the light given by $\abs{R_{u,v}}^2$. This allows judicious phase control to shift error into phase terms where the visual quality is not effected. The second application is spatial. Many applications are only interested in portions of the replay field, allowing error to be moved to the portions of the replay field of lower concern. 
    
    The greatest challenge to utilising SQ is its mathematical complexity. Unless care is taken, expressions for paired movements become quartic and therefore computationally expensive. In the remainder of this work we develop a single example of SQ which uses judicious formulation to avoid quartic solutions and illustrates the power of the SQ approach. We will show that for real-time applications this will allow us to significantly improve on single-frame algorithms such as OSPR and STTM  \cite{cable200453,STTM}.
    
    \section{Soft Sympathetic Quantisation}
    
	In order to demonstrate SQ in action, we present what we are calling soft sympathetic quantisation or SSQ. SSQ is applicable to phase modulated, phase insensitive hologram generation where only the intensity of the replay is of concern and the replay phase is insignificant. This is commonly found in display applications due to the phase insensitivity of the eye. If we adjust pairs of pixels, $H_{x,y}$ and $H_{-x,-y}$, at locations rotationally symmetric around the origin we can write 
    
    \begin{equation}\label{updateSSQ}
    \Delta R_{u,v} = \frac{1}{\sqrt{N_xN_y}}\left[
    \Delta H_{x,y} e^{\left[-2\pi i\left(\frac{ux}{N_x}+\frac{vy}{N_y}\right)\right]}+
    \Delta H_{-x,-y} e^{\left[-2\pi i\left(\frac{-ux}{N_x}+\frac{-vy}{N_y}\right)\right]}
    \right]
    \end{equation}
    
    It can be seen that $\Delta R_{u,v}$ necessarily has angle $0$ or $\pi$ independently of the value of $u$, $v$, $x$ or $y$ provided $\Delta H_{x,y}=\overline{\Delta H_{-x,-y}}$  for all $u$, $v$. Provided these conditions are kept, we can adjust the values of $\Delta H_{x,y}$ and $\Delta H_{-x,-y}$ freely. Note that we here use $\angle$ as the phase operator and here $a\angle b$ represents rotating $a$ by $b$ radians. Additionally, we write the conjugate of variable $x$ as $\overline{x}$.
    
    In simple terms, if we take a pair of pixels symmetric around the origin and modify them so that the change in one pixel is the complex conjugate of the change of the other we can localise the replay field changes due to quantisation to lie on a single line on the Argand diagram.

    \begin{figure}
    	\centering
    	{\includegraphics[trim={0 0 0 0},width=0.6\linewidth,page=1]{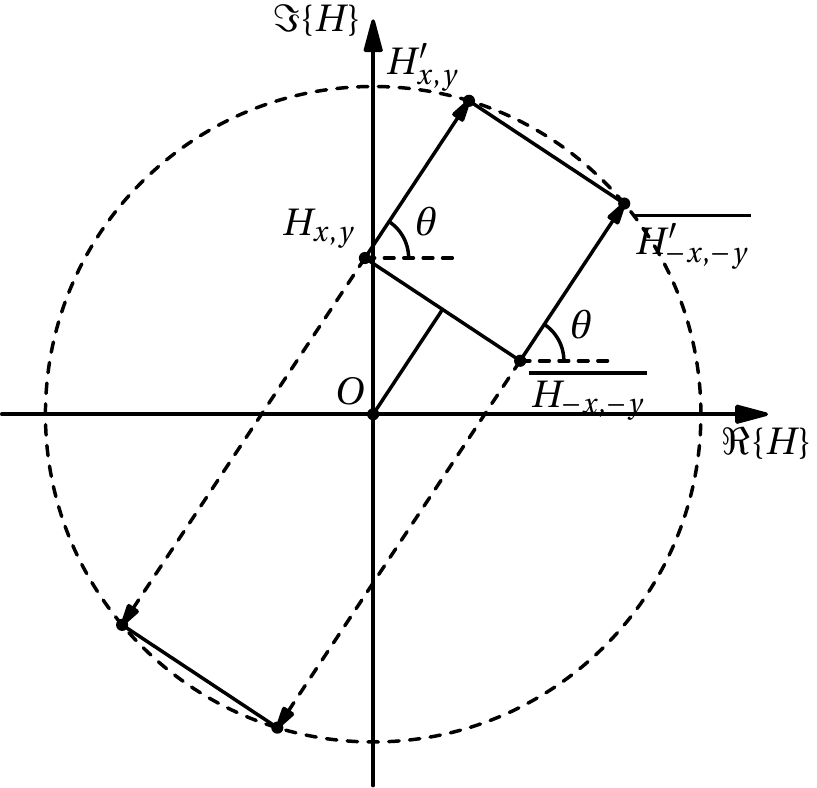}}
    	\caption{Soft sympathetic quantisation for continuous phase devices} 
    	\label{fig:SSQ-Continuous}
    \end{figure}
    
    \subsection{Mathematical Formulation}
    
    The standard form of quantisation which we will term Nearest Neighbour Quantisation (NNQ) is given as
    
    \begin{equation} 
    H'_{x,y}=\exp{(2\pi i \angle H_{x,y}}) , 
		\quad
		H'_{-x,-y}=\exp{(2\pi i \angle H_{-x,-y}})
    \end{equation}
    
    To meet the SSQ constraints, we replace this with a relationship for new pixel values $H'_{x,y}$ and $H'_{-x,-y}$ where
    
    \begin{equation} 
    H'_{x,y}-H_{x,y} = \overline{H'_{-x,-y}} - \overline{H_{-x,-y}} , 
		\quad 
    \abs{H'_{x,y}} = \abs{H'_{-x,-y}} = 1
    \end{equation}
    
    We can represent this geometrically as shown in Figure~\ref{fig:SSQ-Continuous} where the modulation problem becomes the one of transforming the chord between $H_{x,y}$ and $H_{-x,-y}$ in order to lie on the circle when $\abs{H_{x,y}-H_{-x,-y}}\le\nicefrac{2}{\sqrt{N_xN_y}} $ or to lie through the origin in the case $\abs{H_{x,y}-H_{-x,-y}}>\nicefrac{2}{\sqrt{N_xN_y}}$.
    
    It can be shown using the intersecting chords theorem that we can choose vectors $c$ and $m$
    
    \begin{equation}
    c = \frac{H_{x,y}+\overline{H_{-x,-y}}}{2} \nonumber
    \end{equation}
    \begin{equation} \label{eqn:alg1}
    m=\pm   i(H_{x,y}-\overline{H_{-x,-y}})\sqrt{\frac{1}{\abs{H_{x,y}-\overline{H_{-x,-y}}}^2}-\frac{1}{4}}
    \end{equation}
    
    to give
    
    \begin{equation}  \label{eqn:alg2}
    H'_{x,y}=H_{x,y} + m - c, \quad
    H'_{-x,-y}=H_{-x,-y} + \overline{m} - c
    \end{equation}
    
    This approach can be executed in parallel and is negligible in execution time when compared to the FFT element.
    
    Note also, that this formulation ceases to work for $\abs{H_{x,y}-\overline{H_{-x,-y}}}>\nicefrac{2}{\sqrt{N_xN_y}}$ as there is no longer a way of moving the pixel pairs to the circle while still satisfying the constraints. In these cases the points are adjusted so that the point equidistant between them lie at the origin. For the test images we used this occurred less than $0.1\%$ of the time.
    
    \subsection{Phase Randomisation}
    
    Before discussing performance a digression is made to talk about \textit{phase randomisation}. Typically for phase insensitive holograms, the seed image used for the algorithm has a uniformly randomised phase profile. This reduces edge enhancement and serves to smooth the spectral profile. For algorithms like Gerchberg-Saxton (GS) \cite{gerchberg1972practical}, this is only significant for the first few iterations and makes little to no difference to convergent behaviour. For single-iteration algorithms like OSPR and STTM phase randomisation is more important as there is no iterative convergence process.
    
\begin{figure}
    	\centering
    	{\includegraphics[trim={0 0 0 0},width=\linewidth,page=1]{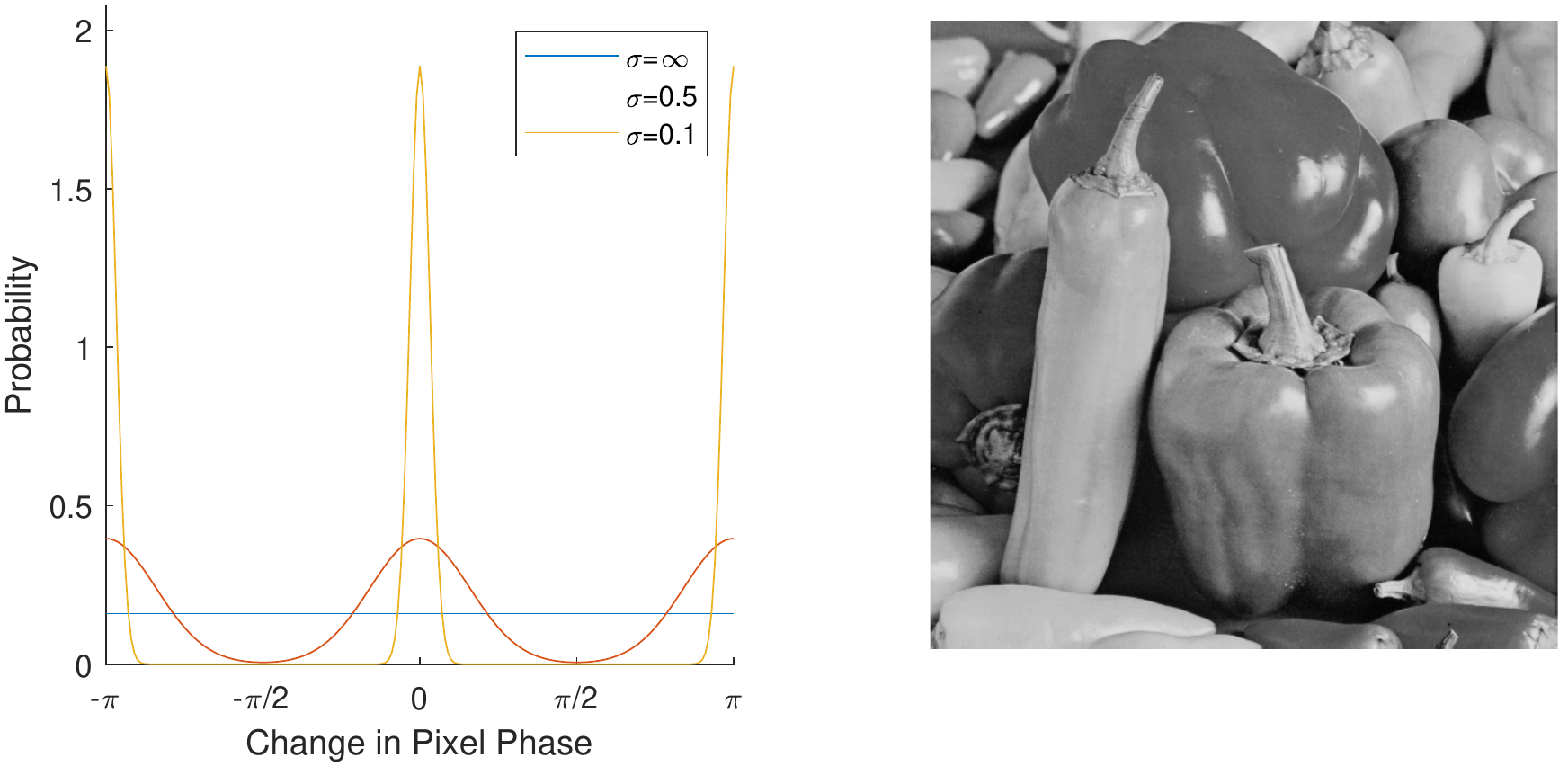}}
    	\caption{Probability densities for phase randomisation (left) and  \textit{Peppers} test image (right).} 
    	\label{fig:sympathetic_009}
    \end{figure}

    For SSQ to work, however, we require the seed phases to lie near to or on a single axis on the Argand diagram. In doing so it is ensured that any errors introduced during quantisation lie perpendicular to this axis, i.e. along the azimuthal direction of the Argand diagram, corresponding to phase errors to which the eye is not sensitive. Instead of the more traditional uniformly distributed seed phase, we use a narrow band phase randomisation approach as shown in Figure~\ref{fig:sympathetic_009} (left) which shows the dual von Mises distribution used.
    
    \subsection{Behaviour}
    
\begin{figure}
    	\centering
    	{\includegraphics[trim={0 0 0 0},width=\linewidth,page=1]{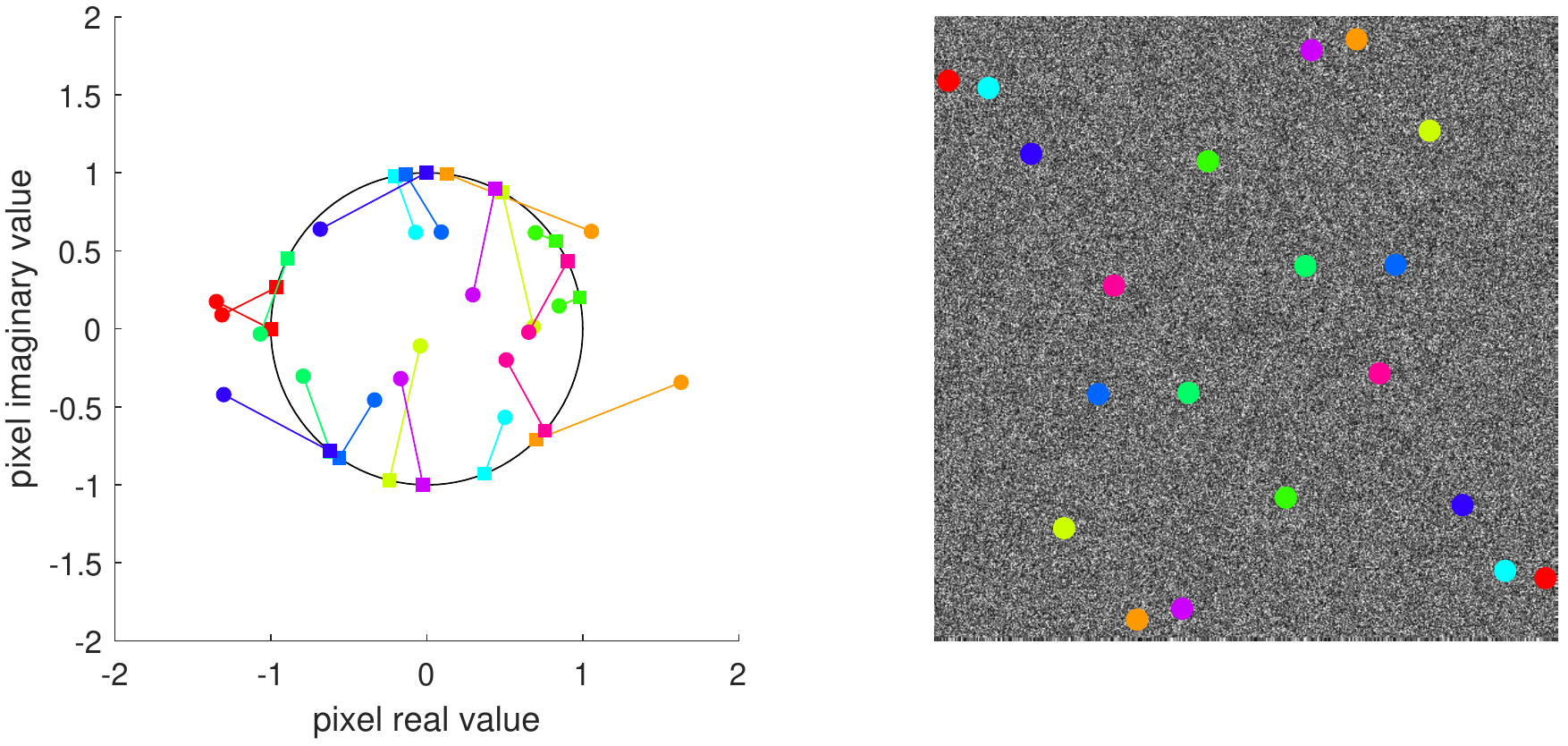}}
    	\caption{Change of 10 randomly selected pixel pairs (left) for the \textit{Peppers} test image showing the movement of the pixel values with starting points shown as circles and end points as squares. The pixel locations on the generated hologram are shown right.} 
    	\label{fig:sympathetic_003b}
    \end{figure}

    In order to better visually understand the change in pixels, we take a selection 10 random pairs of hologram pixels shown in Figure~\ref{fig:sympathetic_003b} (right). We use the relationships in Eqs.~\ref{eqn:alg1}-\ref{eqn:alg2} to quantise the pixels. Figure~\ref{fig:sympathetic_003b} (left) shows the starting values as circles and the final values as squares.
    
     %
    
    \subsection{Algorithm}

    \subsubsection{Single Iteration}
    
\begin{figure}
		\centering
		{\includegraphics[trim={0 0 0 0},width=\linewidth,page=1]{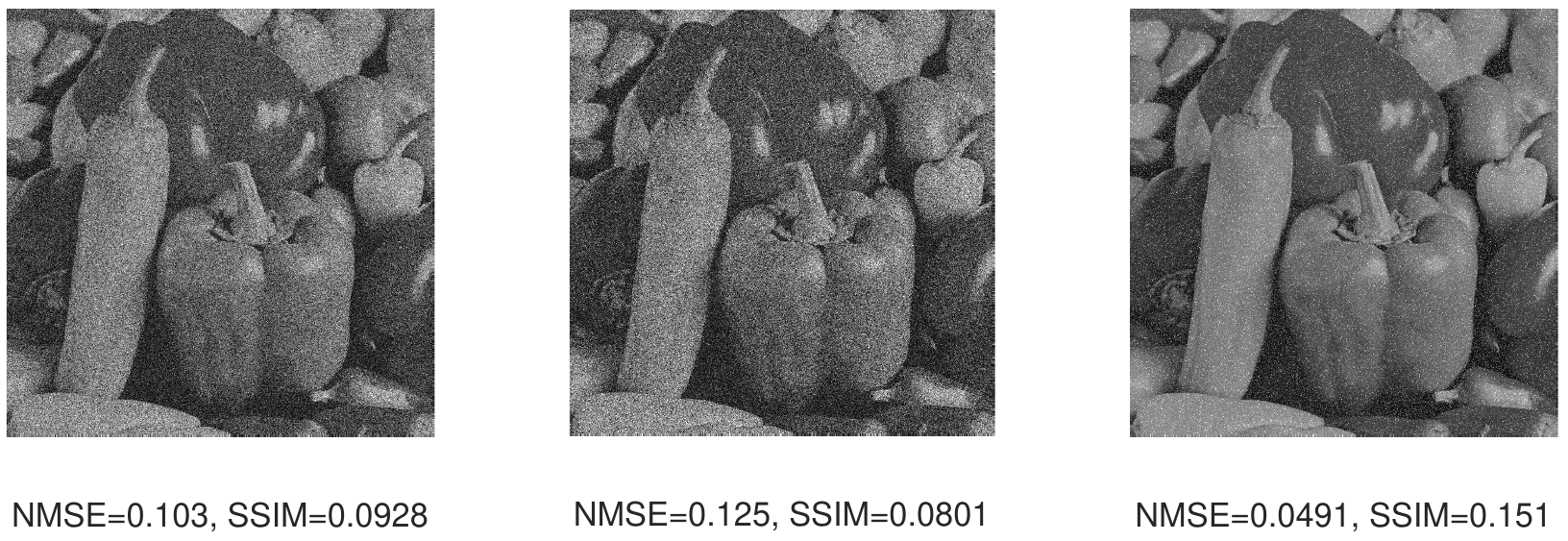}}
		\caption{Comparison of the initial inverse transform with full phase randomisation with NNQ (left) narrow band ($\sigma=0.05$) randomisation with NNQ (centre) and narrow band ($\sigma=0.05$) randomisation with SSQ (right). The SSIM measurements assume a dynamic range equal to 1.  The SLM is assumed to have $256$ levels.} 
		\label{fig:sympathetic_004b}
	\end{figure}

    Figure~\ref{fig:sympathetic_004b} shows holograms produced using full phase randomisation with NNQ (left), narrow band ($\sigma=0.05$) randomisation with NNQ (centre) and narrow band ($\sigma=0.05$) randomisation with SSQ (right).

	The mean squared error (MSE) is calculated according to 
	
	\begin{equation} \label{mse}
	Error(T,R) = \frac{1}{N_x N_y}\sum_{x=0}^{x=N_x-1}\sum_{y=0}^{y=N_y-1} \left(R_{u,v}\overline{R_{u,v}} - T_{u,v}\overline{T_{u,v}}\right)^2 
	\end{equation}
	
	and the structural similarity index (SSIM) is calculated according to 
    
    \begin{equation}\label{ssim}
    SSIM(T,R) = \underbrace{\frac{\left(2\mu_T\mu_R+c_1\right)} {\left(\mu_T^2+\mu_R^2+c_1\right)}}_{S_1}\underbrace{\frac{\left(2\sigma_{TR}+c_2\right)} {\left(\sigma_T^2+\sigma_R^2+c_2\right)}}_{S_2}
    \end{equation}
    
    where $\mu_T$ and $\mu_R$ are the window means; $\sigma_T$ and $\sigma_R$ are the window variances; $\sigma_{TR}$ is the covariance of the two window and $c_1$ and $c_2$ are functions of pixel dynamic range, $L$, where $c_1=(k_1L)^2$ and $c_2=(k_2L)^2$. $k_1$ and $k_2$ are taken as $0.01$ and $0.03$ respectively \cite{wang2004image}.
    

	Case (b) is unlikely to be used in a real-world system but is included to highlight the competing factors. Moving from full randomisation to narrow band randomisation accounts for the decrease in quality between (a) and (b). This is more than compensated for by the addition of SSQ between (b) and (c). Similar results are seen when the algorithm is applied to other standard test images such as \textit{Peppers} and \textit{Camera Man}.
	
	\begin{figure}
		\centering
		{\includegraphics[trim={0 0 0 0},width=\linewidth,page=1]{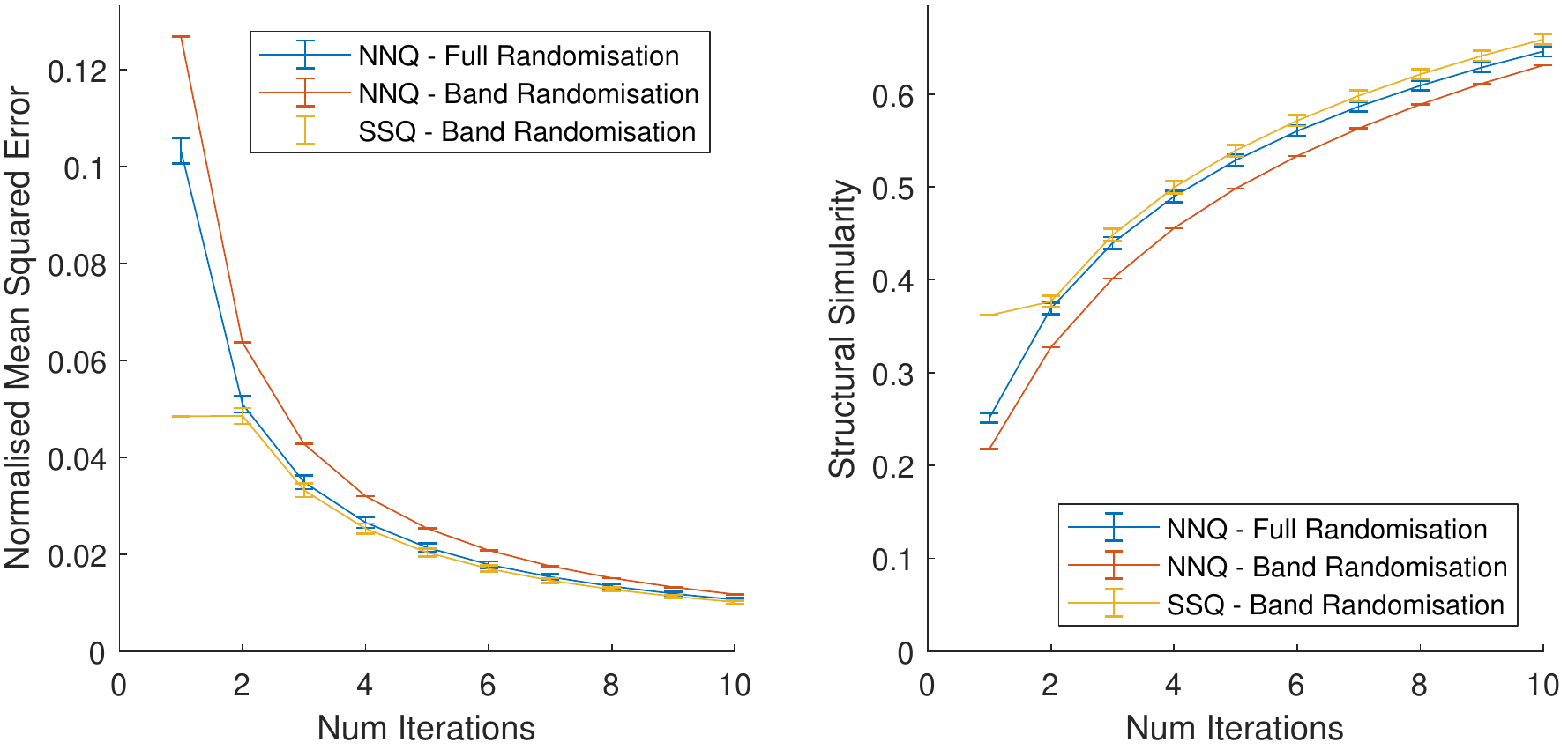}}
		\caption{Convergence of the GS algorithm using SSQ compared to NNQ with full and narrow band ($\sigma=0.05$) randomisation. MSE is shown left with SSIM right. The SSIM measurements assume a dynamic range equal to 1. The SLM is assumed to have $256$ levels.} 
		\label{fig:sympathetic_004d}
	\end{figure}

    \subsubsection{Multiple Iterations}

    SSQ fails to continue working when applied to iterative algorithms such as GS. As shown in Figure~\ref{fig:sympathetic_004d}, the first iteration offers significant performance benefits but the advantages disappear after the first iteration. The reason for this is that only the first iteration is done with a narrow band randomised phase distribution. Once phase distribution is more varied, SSQ becomes detrimental and convergence is worse than the NNQ case. 
    
    This suggests that SSQ is applicable primarily to single iteration approaches such as OSPR and STTM where individual hologram quality is sacrificed in favour of faster generation speed for real-time displays. Here SSQ shows strong improvements in both MSE and SSIM. 
    
	\subsection{Choice of $\sigma$}
	
\begin{figure}
    	\centering
    	{\includegraphics[trim={0 0 0 0},width=\linewidth,page=1]{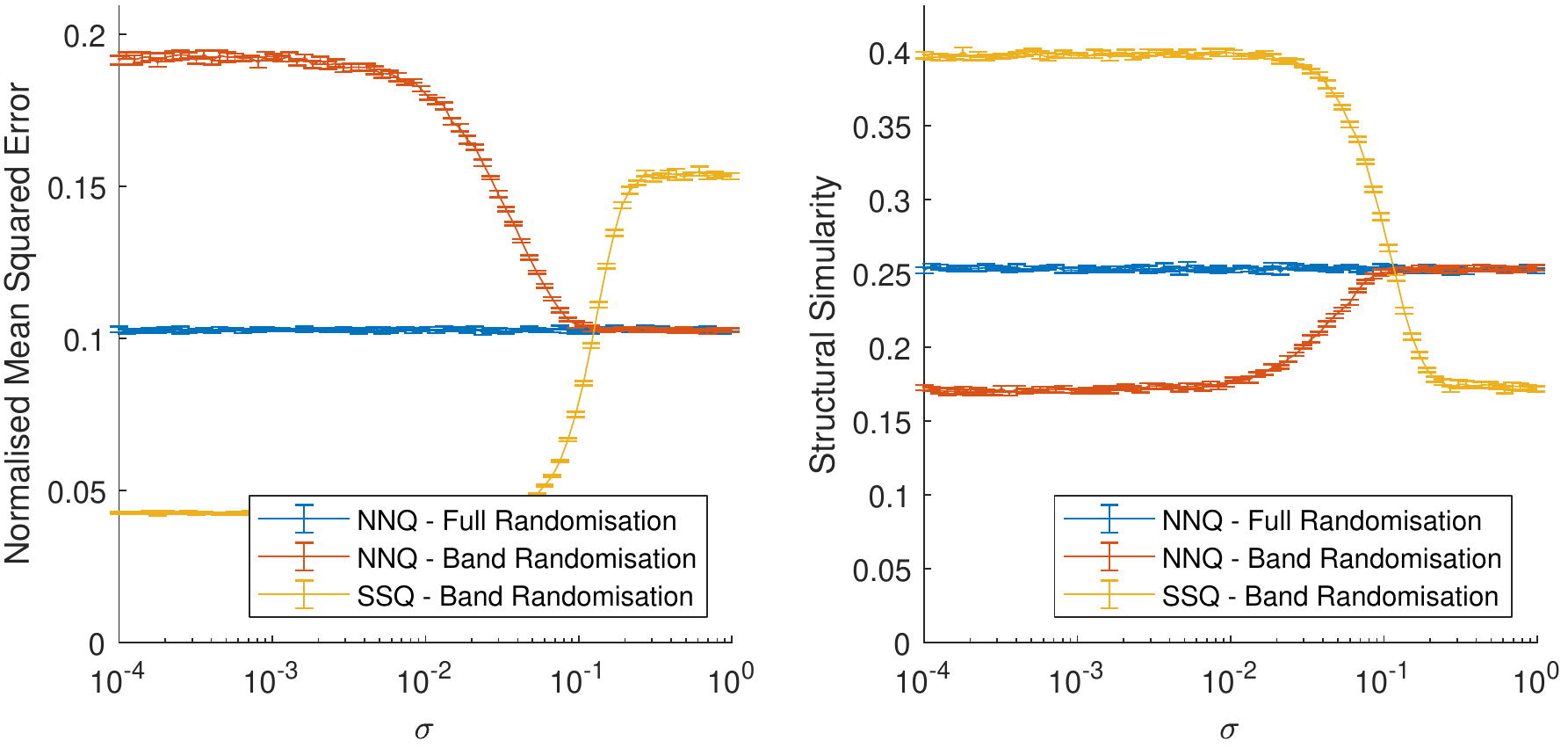}}
    	\caption{Comparison of SSQ performance compared to NNQ with full and banded randomisation against band width. MSE is shown left with SSIM right. The SSIM measurements assume a dynamic range equal to 1. The SLM is assumed to have $256$ levels. Values are taken as being the mean of $20$ independent runs with error bars showing two standard deviations.} 
    	\label{fig:sympathetic_004e}
    \end{figure}

	This prompts the question, what value of sigma should be chosen? Too high and the initial replay phase is no longer sufficiently uniform, too low and edge enhancement effects may well begin to dominate. Figure~\ref{fig:sympathetic_004e} suggests that $\sigma$ can be reduced to near zero.
    
    \subsection{Fresnel Diffraction}
    
    The discussion so far has focussed on Fourier or Fraunhofer holograms. Fresnel holograms can be represented in a similar manner with the addition of a quadratic phase term
    
    \begin{equation} 
    R_{u,v} = \underset{\scriptscriptstyle \text{Fresnel}}{\mathcal{F}}\{H_{x,y}\} = \underset{\scriptscriptstyle \text{Fraunhofer}}{\mathcal{F}}\{H_{x,y}e^{\frac{i \pi}{\lambda z}(x^2 + y^2)}\}
    \end{equation}
    
    where $\lambda$ is the illumination wavelength. Fortunately the rotational symmetry of SSQ means that Eq. \ref{updateSSQ} still applies and by extension Eqs.~\ref{eqn:alg1}-\ref{eqn:alg2}. In our tests we found similar performance and quality  gains in the Fresnel region to those observed in the Fourier region.
    
    \subsection{Quantisation Levels} \label{qlevels}

\begin{figure}
        \centering
        {\includegraphics[trim={0 0 0 0},width=\linewidth,page=1]{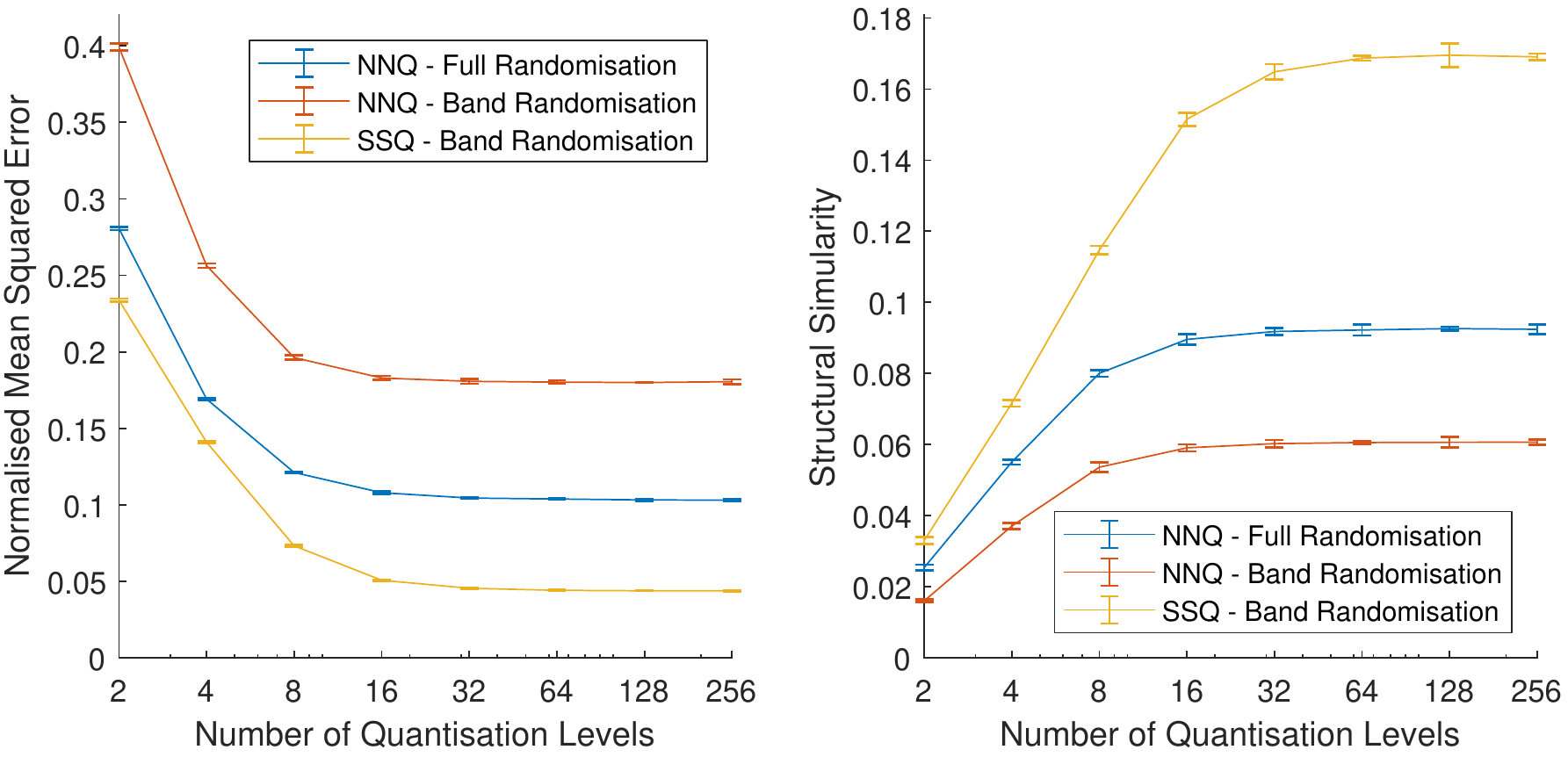}}
        \caption{Comparison of SSQ performance compared to NNQ with full and banded randomisation against number of modulation levels. MSE is shown left with SSIM right. The SSIM measurements assume a dynamic range equal to 1. The narrow band randomisation is taken with $\sigma=0.01$. Values are taken as being the mean of $20$ independent runs with error bars showing two standard deviations.} 
        \label{fig:sympathetic_004f}
    \end{figure}

    All the results presented so far have been for the case of SLMs with 256 quantisation levels and it is worth investigating the case with lower numbers of quantisation levels. Figure~\ref{fig:sympathetic_004f} shows a comparison of MSE and SSIM against number of quantisation levels for the first iteration.

    This shows that SSQ offers the greatest performance improvements for higher numbers of quantisation levels but still offers performance benefits for low numbers of modulation levels.

	\subsection{Applications and Limitations}
	
	For single frame approaches such as OSPR and STTM, SSQ offers the potential to significantly improve both MSE and SSIM. For the example images given, MSE was reduced to under $50\%$ of more traditional approaches while SSIM saw a greater than $50\%$ improvement. This is, unfortunately, limited to only the first iteration of the algorithm. 
    
    A number of time-multiplexed algorithms, such as OSPR or STTM only operate in single frame contexts, time averaging many low quality frames. Here speed of generation is paramount and here SSQ offers significant performance benefits.
	
	The computational overhead of SSQ is low with our implementation spending more than 98\% of runtime on FFT calculation and SSQ requiring less than 0.2\% additional computational overhead. Mathematical complexity is also straight forward with Eqs.~\ref{eqn:alg1}-\ref{eqn:alg2} only requiring simple algebraic manipulation. 

    \section{Conclusions}
    
    This paper has set out to do two things. Firstly to introduce an alternative approach to hologram quantisation and secondly to present a simple example of this in action. 
    
    This work has presented an approach for hologram quantisation called sympathetic quantisation. SQ uses the mathematical formulation of the Fourier transform to adjust pairs of pixel simultaneously during hologram quantisation. This paired movement allows for greater control of the resultant error in the replay field and by extension image quality. By using geometric approaches we are able to avoid the quartic relationships that similar problems often degenerate to.
    
    Significant work is still required to explore alternative formulations in hologram generation. The ability to control the location of replay field error is an exciting opportunity for hologram designers and it is anticipated that this will prove profitable for future study. For example, in fibre mode generation \cite{Carpenter2010} both the amplitude and phase of the replay field are controlled but only for a small central portion. It is anticipated that SQ would allow for a quantisation technique that localised quantisation error to regions outside of the region of interest.
    
    A single example of SQ, soft sympathetic quantisation, has been presented which uses a simple relationship in Eqs.~\ref{eqn:alg1}-\ref{eqn:alg2} to update pairs of hologram pixels located symmetrically around the origin in a manner that moves replay error into phase rather than intensity. For the example images given MSE was reduced to under $50\%$ when compared to traditional NNQ while SSIM saw a greater than $50\%$ improvement. For time-multiplexing single-iteration algorithms such as OSPR and STTM this is a significant performance benefit at negligible cost to performance.
    
    Many questions remain worthy of exploration for SSQ. Firstly, combining SSQ with algorithms more advanced than OSPR or STTM is likely to be beneficial. Secondly, understanding the effect of target image magnitude spectrum on performance is expected to be worthwhile. Perhaps the biggest unanswered question is whether this approach can be extended to greater numbers of pixels. It is anticipated that manipulation of 3 or more pixels may allow for further advanced replay noise control opportunities.
        
    \section*{Acknowledgements}
        
    The authors would like to thank Dr Colin Christopher for initial conversations regarding geometric interpretation of the SQ approach.
        
    \section*{Funding}
    
    The authors would like to thank the Engineering and Physical Sciences Research Council (EP/L016567/1, EP/L015455/1 and EP/T008369/1) for financial support during the period of this research.
    
    \section*{Disclosures}
    
    The authors declare no conflicts of interest.
    
    \bibliography{references}
    
\end{document}